\documentclass{Interspeech}

\interspeechcameraready

\title{The role of audio-visual integration in the time course of phonetic encoding in self-supervised speech models}

\author[affiliation={1}]{Yi}{Wang}
\author[affiliation={1}]{Oli Danyi}{Liu}
\author[affiliation={1}]{Peter}{Bell}

\affiliation{}{University of Edinburgh}{United Kingdom}
\email{Wang.Yi@ed.ac.uk, Oli.Liu@ed.ac.uk, Peter.Bell@ed.ac.uk}
\keywords{audio-visual speech processing, multimodal speech perception, self-supervised learning representation}

\newcommand{\blue}[1]{\textcolor{blue}{#1}}
\usepackage{comment}
\usepackage{multirow}
\usepackage{multicol}

\begin{document}

\maketitle

\begin{abstract}
Human speech perception is multimodal. In natural speech, lip movements can precede corresponding voicing by a non-negligible gap of 100-300 ms, especially for specific consonants, affecting the time course of neural phonetic encoding in human listeners. However, it remains unexplored whether self-supervised learning models, which have been used to simulate audio-visual integration in humans, can capture this asynchronicity between audio and visual cues. We compared AV-HuBERT, an audio-visual model, with audio-only HuBERT, by using linear classifiers to track their phonetic decodability over time. We found that phoneme information becomes available in AV-HuBERT embeddings only about 20 ms before HuBERT, likely due to AV-HuBERT's lower temporal resolution and feature concatenation process. It suggests AV-HuBERT does not adequately capture the temporal dynamics of multimodal speech perception, limiting its suitability for modeling the multimodal speech perception process.
\end{abstract}

\vspace{-0.2cm}
\section{Introduction}
Recent years have seen a rising trend in using artificial neural networks (ANN) to model sensory processes in humans \cite{kanwisher2023using}.
For speech processing, it has been shown that the neural activity in the auditory cortex of human listeners correlates with representations in a recurrent neural network model trained for speech recognition, with correspondence between levels of language processing and different layers within the model \cite{keshishian2025parallel}.
Apart from encoding hierarchy, their findings also revealed striking resemblance between the RNN model and the human brain in terms of the temporal order of linguistic representations. 
This comparison between model representations and brain activities could potentially help us identify the brain area and the time course in which different levels of language processing unfolds.

While various characteristics of speech processing have been examined in ANN models in comparison to human listeners, most existing work has focused on models with only audio input.
In reality, human speech perception process is fundamentally multimodal, involving audio-visual integration, as illustrated by the McGurk effect \cite{mcgurk1976hearing}, in which contradictory audio and visual cues cause a shift in human phoneme perception.
While previous work on multisensory integration has studied how simple motion and spatial cues are integrated \cite{stein2008multisensory}, the integration mechanism of higher-order semantically complex information, such as audio-visual cues in speech, remains unclear.

Some prior work  \cite{ngiam2011multimodal,grasse2024role} have used audio-visual speech recognition models as a tool for investigating audio-visual speech perception.
In contrast to \cite{keshishian2025parallel}, they aimed at answering the developmental questions of brain speech processing,
and trained ASR models under specific conditions to see if such conditions can trigger the development of corresponding characteristics in ANN models.
For example, audio-visual ANNs trained by unsupervised learning proposed by Ngiam et al. \cite{ngiam2011multimodal} did exhibit the McGurk effect. 
Furthermore, a subsequent study \cite{grasse2024role} reported that the McGurk effect could emerge without specific training on the latest audio-visual self-supervised learning (SSL) speech representation model, Audio-visual HuBERT (AV-HuBERT) \cite{avhubert2022learning}.
They thus suggested that audio-visual self-supervised learning (SSL) models could be useful for simulating how humans perform audio-visual integration.

A crucial factor in audio-visual integration during speech perception, overlooked in \cite{grasse2024role}, is the earlier onset of the visual signal relative to the corresponding audio signal.
Statistical analyses indicate that in natural speech production, mouth movements precede the resulting vocalization by approximately 100 to 300 milliseconds in the bilabial and labial consonants they tested \cite{chandrasekaran2009natural}. 
This lead time, known as the ``time-to-voice," reflects a fundamental characteristic of audio-visual speech timing.
Humans have adapted to such asynchronicity, as evidenced by the 
reduction in amplitude and latency of auditory event-related potentials when speech audio is paired with visual stimuli \cite{besle2004bimodal, van2005visual, zeng2024asynchronicity}. 
However, it is unexplored to what extent incorporating visual cues accelerates the decodability of phonemes in ANN representations. 
If an ANN model exhibits little to no difference in the time course of phonetic encoding, the validity of using the model to simulate temporally sensitive aspects of speech perception would be significantly compromised.

In this paper, we investigate whether the incorporation of visual signals affects the time course of phonetic encoding in a self-supervised learning (SSL) speech model, as it does with human listeners. 
To this end, we compare an audio-visual model, AV-HuBERT, with its audio-only counterpart, HuBERT, by examining the window of phonetic decodability \cite{liu2024predictive, gwilliams2022neural}.
Using a set of linear classifiers, we obtain a fine-grained view of when information about a phone becomes available in the model before its onset in the acoustics, with a temporal resolution up to the frame rate of the representations. In this way, we model how much of the temporal dynamics of visual and audio signals in speech can be encoded by the SSL representations.

We find that while visual cues can precede acoustic cues by hundreds of milliseconds, they advance phonetic decodability in AV-HuBERT over HuBERT only by approximately 20 ms.
This limited effect is likely due to AV-HuBERT’s down-sampling on the acoustic features by stacking every four frames of acoustic features to align with each visual feature frame.
This process advances the audio features by an average of approximately 20 ms from their original timing.
Furthermore, by manually introducing greater asynchronicity between visual and acoustic signals when depicting decodability curves, we find that AV-HuBERT’s phonetic encoding is more strongly influenced by acoustic than visual signals. 
Overall, our findings suggest that greater caution is needed when using AV-HuBERT to model temporally sensitive aspects of multimodal speech perception.

\vspace{-0.2cm}
\section{Related works and method}
\label{section: Method}
Our analysis relies on using pre-trained SSL models as feature extractors, and training phoneme classifiers to identify the decodable window of a phone.
\vspace{-0.1cm}
\subsection{HuBERT and Audio-visual HuBERT}
\label{subsection: HuBERT HuBERT}
\vspace{-0.1cm}
We used an audio-only and an audio-visual SSL model as the feature extractors for the corresponding modality. 
We chose the AV-HuBERT model \cite{avhubert2022learning} and its audio-only counterpart HuBERT \cite{hsu2021hubert}, as AV-HuBERT has been the most widely used SSL audio-visual model. Both models were pre-trained by the masked prediction of hidden units method \cite{hsu2021hubert}. For abbreviation, HuBERT and AV-HuBERT will hereinafter be addressed as ``HuBERTs'' if being described together.

AV-HuBERT shares a similar masked prediction pre-training paradigm as HuBERT, but with the following modifications: 
1) involving cross-modal transfer learning in the first pre-training iteration by
using the pseudo labels extracted from acoustic features (MFCCs) as the initial pre-training target;
2) adding a deeper feature extractor ResNet-18 \cite{he2016deep} to encode visual inputs;
3) applying early fusion at feature level by concatenation before feeding the audio and visual features into the transformer structure. More specifically, to align the audio log Mel-filterbank energy feature at 100 Hz to the visual ResNet-18 feature at 25 frames per second, every four audio feature frames are stacked and concatenated with a visual feature. The amount and domain of data being used for pre-training AV-HuBERT and HuBERT are also different.   

Our analysis in Section \ref{section: Experiments} conducts a normalization step on the decodability curve to mitigate the impact from such design differences, and will focus on revealing the characteristic of AV-HuBERT more by comparing the peak time of its decodability curve to the phone onset time. 

In the feature extraction steps of all experiments in this work, we used the publicly available HuBERT BASE \footnote{\label{note1}https://github.com/facebookresearch/fairseq/tree/main/examples/hubert} pre-trained on the 960-hour subset of LibriSpeech,
and the AV-HuBERT BASE model \footnote{\label{note1}https://github.com/facebookresearch/av\_hubert} pre-trained on the 433-hour subset of LRS3,  both model consisting of 12 transformer blocks.
\vspace{-0.2cm}
\subsection{Phonetic decodability window}
\label{subsection: Phonetic decodability window}
Previous works \cite{gwilliams2022neural, khalighinejad2017dynamic} decoded phone identity from brain recordings within a context window of 200-400 ms centered at the phone onset. The analysis was further applied in \cite{liu2024predictive} on audio-only SSL speech representations \cite{oord2018representation} to observe encoded temporal dynamics.

Taking a context window with duration $T$ centered at phone onset [-$\frac{T}{2}$,  $\frac{T}{2}$], it covers $2w + 1 = \frac{T}{sample\_rate}$ frames of speech features $f_{-w:w}$. Given a set of data consists of $N$ phone-level speech sequences $s_{1:N}$ cut from utterances and labelled by phone identity $p_{1:N}$ obtained from forced-alignment, we use $f_{i,j}$ to represent feature in phone-level sequence $s_i$ at time step $j$ relative to the center. 
All frames that share the same time step $j$ relative to the phone onset $F^j$ = $f_{1:N, j}$ can be paired with phone labels $p_{1:N}$ to train a Logistic Regression classifier $C^j$ on phone classification. When testing the classifier $C^j$ on a batch of evaluation speech feature frames at time step $j$, i.e. $f^{eval}_{1:N,j}$, the average classification accuracy $Acc_j$ reflects the ability of decoding the phone identity from the SSL model feature $j$ steps preceding / following the phone onset. 
The curve drawn from predictive accuracy score sequence {$Acc_{-w}$:$Acc_w$} versus the relative time steps in the context window ${-w:w}$ reflect how the temporal dynamics of speech input is captured by the SSL model features. 

We extracted HuBERT and AV-HuBERT features from speech within the context window of each phone, and drew the predictive decoding curve based on each type of features in a 1200 ms time window, which covers 30 frames of AV-HuBERT feature sequence or 60 frames of HuBERT feature sequence, resulting in 30 / 60 classifiers trained to draw an audio / audio-visual HuBERT decodablity curve. 

\vspace{-0.2cm}
\section{Task description}
\label{section: Task description}

\begin{figure*}[t]
\vspace{0cm}
  \centering
  \includegraphics[width=\linewidth]{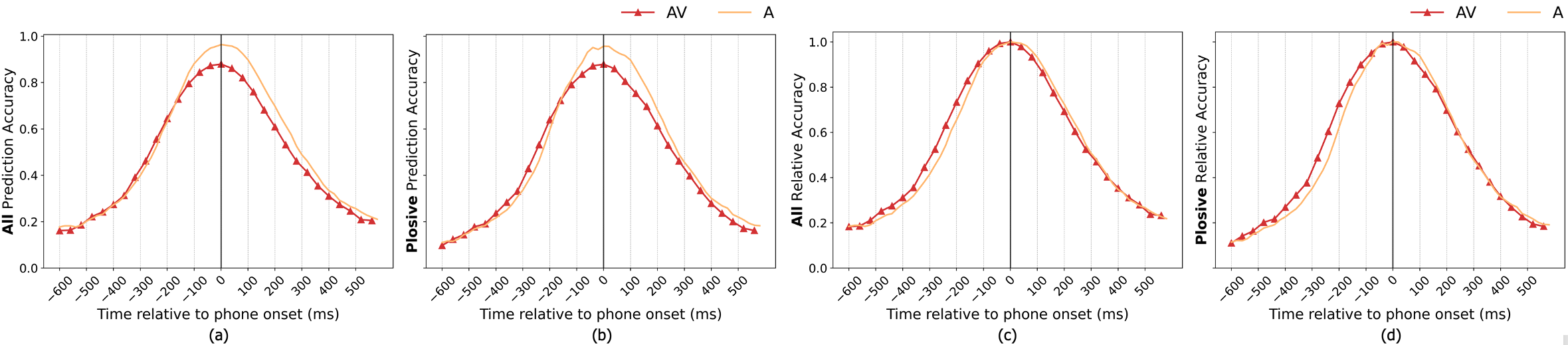}
  \caption{Phonetic decodability curves on Grid data cross-validation experiments, representing (a) absolute prediction accuracy on all phonemes, (b) absolute prediction accuracy on plosive stop consonants, (c) relative accuracy on all phonemes, (d) relative accuracy on plosive stop consonants, with x axis labels representing time (ms) relative to phone onset (0). Plain orange line denotes HuBERT and triangular-marked red lines denotes AV-HuBERT. } 
  \vspace{-0.3cm}
  \label{fig:full context}
\vspace{-0.2cm}
\end{figure*}
\vspace{-0.1cm}

\subsection{Dataset}
\label{subsection: Dataset}
We analyzed the phonetic decodability window on two datasets separately. The phone boundaries and labels for both datasets were obtained through the Montreal forced aligner on English arpabet dictionary \cite{mcauliffe2017montreal, gorman2011prosodylab}.
\\
\textbf{Grid}: The GRID dataset \cite{cooke2006audio} includes speech videos collected in a controlled environment, with 34 speaker sitting at the same spot relative to the camera, front-faced video and noise-free audio being recorded simultaneously. It consists of each speaker pronouncing 1000 sentences with a fixed six-word structure: command(4) color(4) preposition(4) letter(25) digit(10) adverb(4), with the number in parentheses indicating the vocabulary size at the corresponding position. 
\\
\textbf{LRS3}:
The LRS3 dataset \cite{afouras2018lrs3} includes TED and TEDx videos, along with subtitles and word boundaries, and has three sets: pre-train, train-val and test. In calculating the phonetic decodability in context window, we adopted LRS3 train-val split, consisting of 30 hours of speech from 4004 speakers. 

Considering that utterances in Grid has little flexibility in its structure and limited vocabulary, each frame of SSL features may capture context information that covers neighbouring phones, allowing the classifier to distinguish phone identity from context phones with long time away from the target phone.
\vspace{-0.2cm}
\subsection{HuBERT models feature extraction settings}
\label{subsection: HuBERTs settings}
\vspace{-0.1cm}
We conducted three-fold cross-validation on both datasets. Then in actual experiments, we randomly selected $\frac{1}{15}$ samples from both the training and the validation split of the data, further split them into 20 batches during LR training and inference, and averaged the resulting scores across the batches and folds but on each time step separately. Data sampling is due to the limitation of sklearn LR package . 
\vspace{-0.2cm}
\section{Experiments}
\label{section: Experiments}
\begin{figure}[t]
  \centering
    \includegraphics[width=\linewidth]{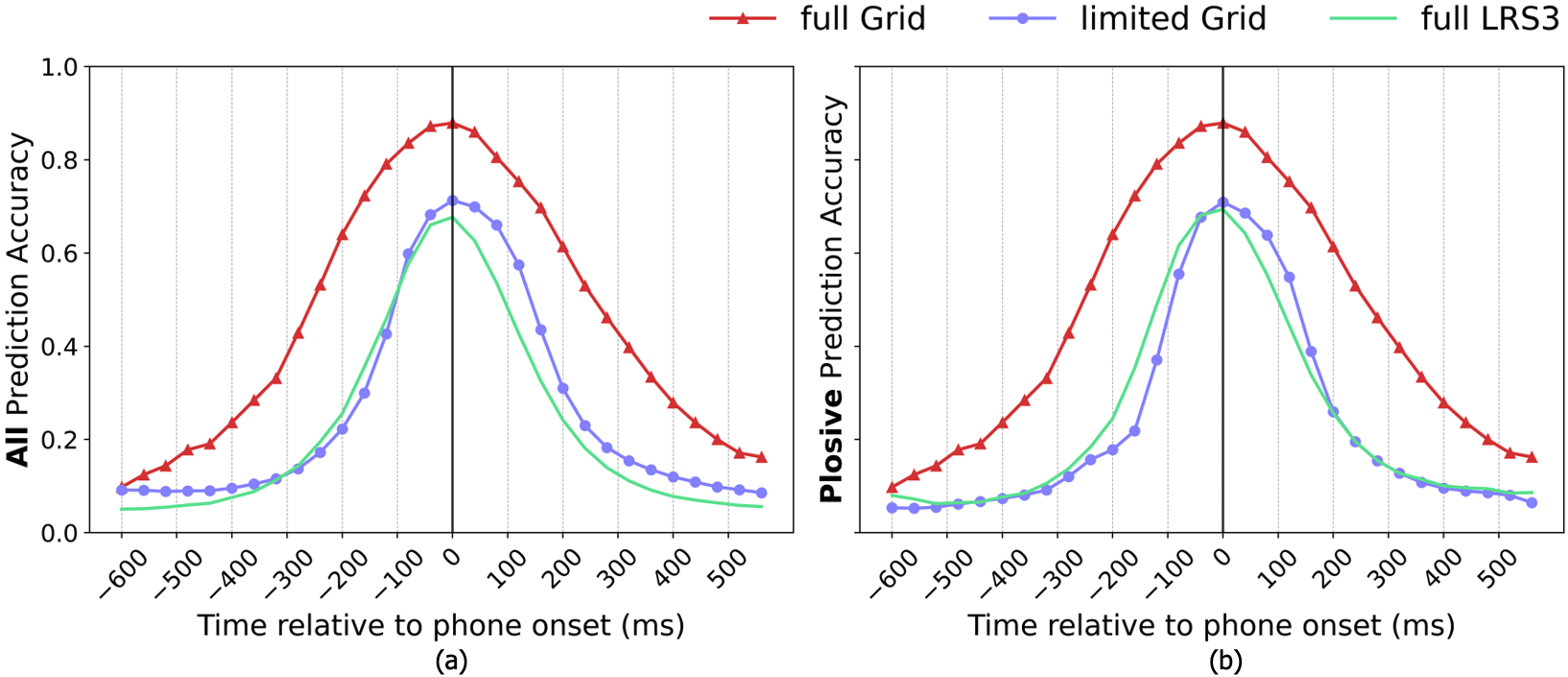}
  \caption{
    Phonetic decodability curves from audio-visual features in the cross-validation experiments, full-context on Grid (full Grid, triangular-marked red lines); limited-context window of 200ms on Grid (limited Grid, circle-marked blue lines) or full-context on LRS3 (full LRS3, plain green lines), plotted on all phonemes (a), or on plosive stop consonants (b). 
  }
  \vspace{-0.3cm}
  \label{fig:LRS3-Grid}
\vspace{-0.2cm}
\end{figure}

\begin{figure}[t]
\vspace{0cm}
  \centering
  \includegraphics[width=\linewidth]{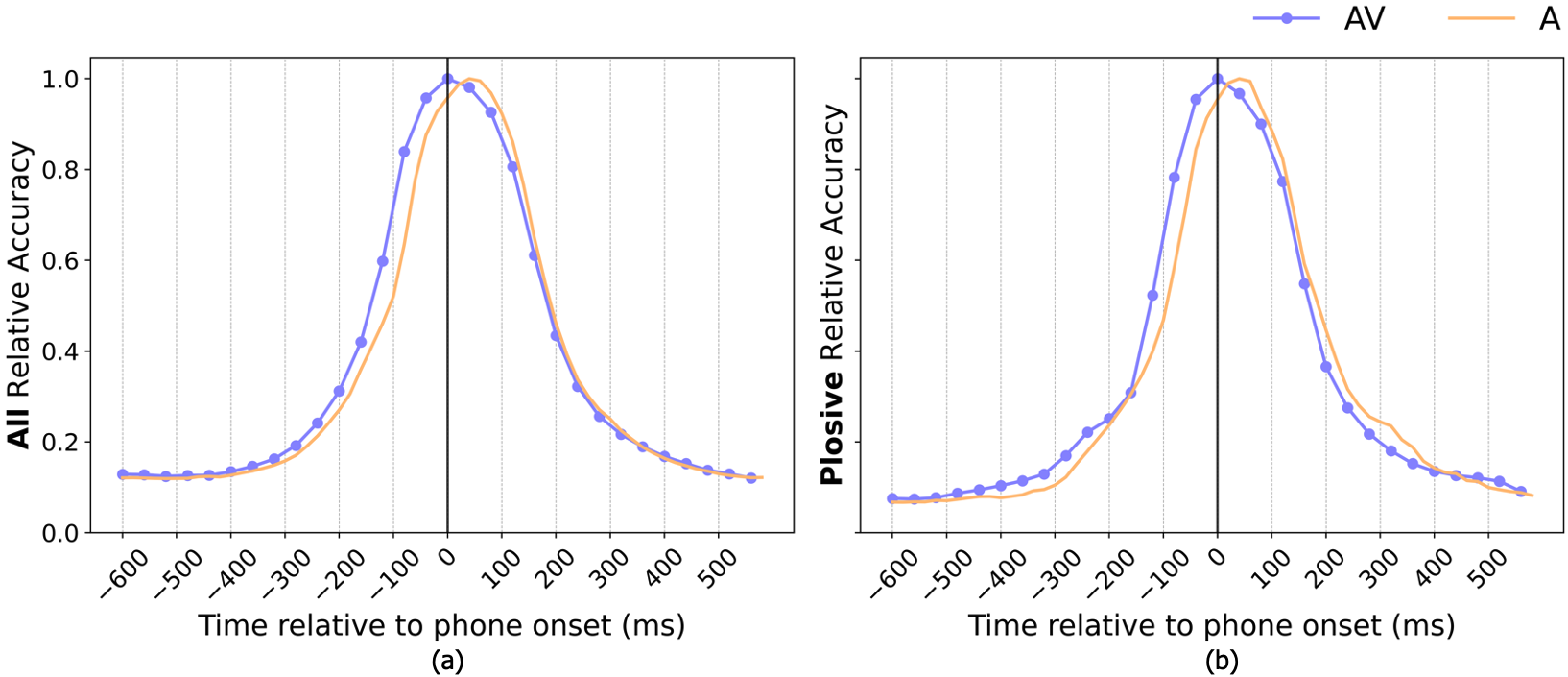}
  \caption{Relative prediction accuracy curves on Grid data based on limited-context window, (a) all phoneme and (b) plosive stops . Circle-marked blue lines for AV-HuBERT. Plain orange lines for HuBERT} 
\vspace{-0.3cm}
  \label{fig:limited_conext}
\vspace{-0.3cm}
\end{figure}

\begin{figure*}[t]
\vspace{0cm}
  \centering
  \includegraphics[width=\linewidth]{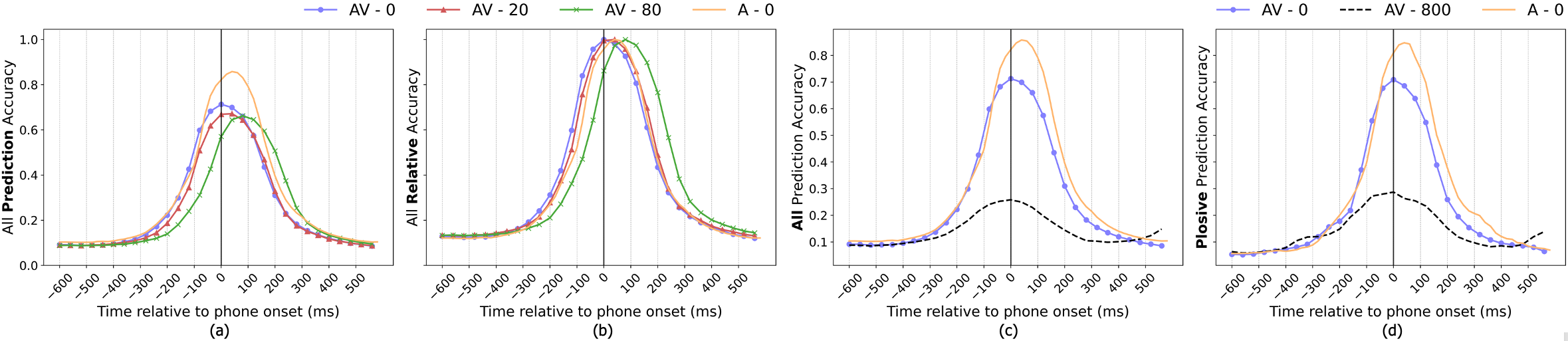}
  \caption{Phonetic decodability curves on asynchronous audio-visual signals from Grid data: (a) absolute prediction accuracy on all phonemes (b) relative accuracy on all phonemes,
  (c) absolute prediction accuracy on all phonemes (d) absolute prediction accuracy on plosive stops. Subfigure (a) and (b) present AV-HuBERT curves with audio signals delayed by 0, 20 and 80 milliseconds (labelled by AV-0/20/80), respectively, and a HuBERT curve on synchronized audio-visual signals (labelled by A-0). Subfigure (c) and (d) include AV-HuBERT and HuBERT curves on synchronized audio-visual signals (labelled by AV/A-0), and an AV-HuBERT curve  with audio signals delayed by 800 ms (labelled by AV-800).} 
\vspace{-0.3cm}
  \label{fig:grid_async}
\vspace{-0.3cm}
\end{figure*}

\subsection{Full context window}
\label{subsection: Full window}
We initially experimented with HuBERTs features extracted given full utterance as inputs, which potentially allows them to encode information from other phones in the utterance. Layer 9 of HuBERT / AV-HuBERT is selected for feature extraction, because it has relatively high correlation to phonetic information among layers \cite{hsu2021hubert, avhubert2022learning}.
Apart from drawing decodability curves averaging samples from all phoneme categories, we also drew from only plosive stops, due to previous work \cite{zeng2024asynchronicity} suggested more prominent time-to-voice on such consonants.

Fig. 1 (a) and (b) present phonetic decodability curves, which show prediction accuracy against time relative to the phone onset, for all phonemes (Fig. 1a) and for plosive stop consonants (Fig. 1b), respectively, plotted with data from Grid. 
HuBERT curves are mildly shifted towards right than centered at the phone onset time, compared to the AV-HuBERT curves (Fig. 1a \& 1b), but there is no significant sign of AV-HuBERT decodability window preceding that of HuBERT. Then we normalized each absolute accuracy curve by its maximum value, as shown in Fig. 1 (c) and (d), trying to mitigate the interference on inspecting the timing of the window from the curves' amplitude and focus only on the decodable timing.
Considering the relative decodability curves (Fig. 1c \& 1d), AV-HuBERT features become decodable approximately 20 ms earlier than HuBERT, which is too small to be an evidence of visual cues accelerating the decodability of phonemes. 

Meanwhile, the normalized AV-HuBERT and HuBERT curves (Fig. 1c \& 1d) show similar shape and width of decodability window, making further comparisons easier.

\vspace{-0.2cm}
\subsection{Limited context window}
\label{subsection: Limited context window}
We further attempted to reduce the impacts from contextual phonetic information as mentioned in Section \ref{subsection: Dataset}, by depicting decodability curve on LRS3 dataset or still on Grid data but with limited context (LC) provided to SSL models. A context window of  200 ms, centered on the corresponding time step of each feature frame, is enforced, with speech signals outside this window replaced by zero padding. Fig. \ref{fig:LRS3-Grid} presents the resulting decodability windows drawn from absolute prediction accuracy on Grid with limited context (circle-marked blue lines) compared to that on full-context Grid data (triangular-marked red lines) and full-context LRS3 data (plain green lines). 
From comparing the Grid LC window curves to the Grid full context curve, we observed that the approach of limiting contexts leads to a reduced performance and narrower decodability windows, making it approximate the size and shape of the LRS3 curve. Taking the LRS3 curve as a reference, we regard applying context window at 200 ms as an effective way to reduce impacts from overwhelming phonetic contexts. Meanwhile, the LRS3 curves precede the Grid LC curves, indicating influences from data domain on the decodability curves.
Potential reasons include the asynchronicity between audio and video recordings, the precision of forced-alignment results, etc. So we will consistently use Grid data with limited contexts in following experiments, considering that LRS3 consists of more noisy data.

We then presented the LC feature generated decodability curves in Figure \ref{fig:limited_conext}, which shows a high similarity between the curves of AV-HuBERT and HuBERT. The AV-HuBERT feature becomes decodable slightly earlier than HuBERT and reaches the most decodable point 20 ms in advance. 
We hypothesized that such a 20 ms time gap is not because AV-HuBERT managed to capture the earlier onset of visual signals. AV-HuBERT model downsamples audio inputs by stacking 4 frames of the audio low-level features before concatenating them with one visual feature frame, so that their temporal resolutions are consistent for joint processing.
This downsampling step blurs the temporal resolution in audio signals, which we suspected causes the 20 ms gap.
More specifically, suppose the probability for a phone onset to fall in each of the four frames being equal, the audio feature corresponding to exactly the phone onset time will be advanced by an expectation of 20 ms. HuBERT can also blur the temporal resolution in audio signals, but its shorter frame window length (20 ms) will lead to shorter time shifts being expected. 

\vspace{-0.3cm}
\subsection{Asynchronicity simulation}
\label{subsection: Asynchronicity simulation}
\vspace{-0.1cm}
The reasoning on AV-HuBERT's relatively decodable window advancing HuBERT by 20 ms, as mentioned in Section \ref{subsection: Limited context window}, relies on a hypothesis, that the timing of  AV-HuBERT's decodability curve is dominated by its audio input. We then manipulated audio signals for AV-HuBERT, delaying it by time $t_d$ compared to the visual inputs and phoneme labels. 
Fig. \ref{fig:grid_async} presents the AV-HuBERT decodability curves using manipulated audio-visual inputs with $t_d \in \{20, 80\}$ ms, compared with the original curve.
The peak times relative to phone onset measured on AV-HuBERT curves with audio signal delayed by $t_d = 0, 20, 80$ are $0, 40, 80$ ms, respectively (Fig. \ref{fig:grid_async}(b), the circle-marked blue, the triangular-marked red, and the cross-marked green lines). Such delays in peak are aligned with the delay time durations in audio signals, considering a delay of 20 ms may be rounded by the 40-millisecond time resolution of AV-HuBERT. The same phenomenon is observed on the plosive stop curves, whose plots are omitted here. The curve with 20 ms audio delay almost overlaps with that of audio-only HuBERT in Fig. \ref{fig:grid_async}(b), (triangular-marked red line vs. orange plain line), supporting our reasoning on AV-HuBERT features' temporal information being dominated by the audio modality while its temporal resolution being perturbed due to downsampling.

Another reason may explain why the decodability curve delays are aligned with the audio signal delays. SSL models may have learned to be robust to asynchronous audio-visual inputs by backing-off on the more informative audio modality.
To rule out such possibility, we applied an extreme delay $t_d = 800$ ms on audio signals, and compared its curve with the original AV-HuBERT curves, as in Fig. \ref{fig:grid_async} (c)(d). 
800 ms of asynchronicity disables the 200-millisecond-long LC window to encode most of the relevant audio signals and ruins models' potential backing-off mechanism, so the decoding primarily relies on the visual input encoding. For both all / plosive stop phonemes, the  black dotted curves in Fig. \ref{fig:grid_async}, which represent the 800 ms asynchronicity, are centered at 0 ms. There is a low bump-up in the plosive curve at around -350 ms, but having too low accuracy to be addressed as 'decodable'. If AV-HuBERT provides adequate encoding in the visual modality based on the actual temporal dynamics of visual signals, we expect the peak to locate prominently earlier then the acoustic phone onset time. We observed that AV-HuBERT provided a temporally skewed visual encoding that follows the temporal dynamics of audio signals. This is likely because the initialization of AV-HuBERT's pseudo labels, based on MFCC features, forced the encoders of both audio and visual modalities to learn acoustic-based targets.

\begin{table}[th]
  \caption{The time of decodability curves' peak relative to phone onset, depicted on representations from different layers of HuBERT/AV-HuBERT models, for all phonemes or for plosive stops. L3/6/9/12 denotes the layer 3/6/9/12 in the corresponding SSL models. 800 Async denotes curves drawn with audio inputs delayed by 800 ms}
  \label{tab:peak}
  \centering
\scalebox{0.8}{
\begin{tabular}{ccccccc}
\noalign{\hrule height 0.8pt}
No. & Model Type                                                                         & \begin{tabular}[c]{@{}c@{}}Phoneme \\ Category\end{tabular} & L3  & L6  & L9  & L12 \\ \noalign{\hrule height 0.5pt}
1   & \multirow{2}{*}{HuBERT}                                                            & all                                                         & 20 & 20 & 20 & 20 \\ \cline{3-7} 
2   &                                                                                    & plosive                                                     & 20 & 20 & 20 & 0  \\ \cline{2-7} 
3   & \multirow{2}{*}{AV-HuBERT}                                                         & all                                                         & 0  & 0  & 0  & 0  \\ \cline{3-7} 
4   &                                                                                    & plosive                                                     & 0  & 0  & 0  & 0  \\ \cline{2-7} 
5   & \multirow{2}{*}{\begin{tabular}[c]{@{}c@{}}AV-HuBERT\\ (800ms Async)\end{tabular}} & all                                                         & 0  & 0  & 0  & 0  \\ \cline{3-7} 
6   &                                                                                    & plosive                                                     & 0  & 0  & 0  & 0  \\ \noalign{\hrule height 0.8pt}
\end{tabular}
}
\vspace{-0.4cm}
\end{table}

\vspace{-0.2cm}
\subsection{decodability window from different layers}
\label{subsection: Asynchronicity layers}
\vspace{-0.1cm}
All experiments conducted in Section \ref{subsection: Full window} to  Section \ref{subsection: Asynchronicity simulation} used features from the 9th layer of HuBERTs models. We further validated that the lack of encoding visual speech temporal dynamics is not only limited to layer 9 features. We regard the time with the highest accuracy on a decodable curve as its peak. The timing of decodability curve's peak drawn on HuBERT/AV-HuBERT layer 3/6/9/12 features are listed in Table \ref{tab:peak}. 
Peak time number 5 and 6 were generated from AV-HuBERT features using asynchronous inputs with 800 ms lags in audio, approximating the encoding of visual speech when audio is absent. Tab. \ref{tab:peak} showed the peak time from other layers throughout HuBERT/AV-HuBERT models approximately align with those from layer 9 described in previous subsections. It confirms that other layers in AV-HuBERT can hardly capture the earlier onset of visual cues either.

\vspace{-0.2cm}
\section{Conclusion}
\vspace{-0.1cm}
In this paper, we investigated how audio-visual SSL models capture the natural asynchronicity between audio and visual cues in speech, by using a sequence of linear classifier to decode phone identity from each frame of AV-HuBERT/HuBERT model features in a context window centered at phone onset. We found that AV-HuBERT's encoding of speech temporal dynamics is dominated by its audio input. The time course of its visual encoding is overwhelmingly aligned to its audio encoding, leading to the natural asynchronicity in speech being overlooked. We believe this is due to the cross-modal transfer learning in AV-HuBERT pre-training initialization. We conclude that great caution is needed if using AV-HuBERT to model multimodal speech perception. Furthermore, all transformer models used in temporal sensitive modelling of human speech and language processing should be used with extensive caution, as the temporal correlation in their features can be manipulated by the design of their training targets.

\section{Acknowledgements}
This work was supported by the UKRI Centre for Doctoral Training (CDT) in Natural Language Processing, funded by the UKRI grant EP/S022481/1 and the University of Edinburgh. We would like to thank Biao Zeng from University of South Wales and Hao Tang, Sharon Goldwater from ILCC, University of Edinburgh for useful discussion. We would like to thank Ondrej Klejch for emergency technical support on forced alignment, and Scott Ye for suggestions on plotting.


\bibliographystyle{IEEEtran}
\bibliography{mybib}

\end{document}